\documentclass[aps%
  ,prc%
  ,10pt%
  ,twocolumn
  ,twoside%
  ,superscriptaddress%
  ,nofootinbib%
  ,showpacs%
  ,showkeys%
  ,amsfonts%
  ,amssymb%
  ,amsmath%
  ]{revtex4-1}

\usepackage{graphicx}
\usepackage{bm}
\usepackage{color}
\usepackage{hyperref}
\usepackage{amssymb,amsfonts,amsmath}
\newcommand{\bvec}[1]{\ensuremath{\boldsymbol{#1}}}

\begin{document}

\title{Antiproton-nucleus reactions at intermediate energies}

\author{A.B. Larionov}
\email{E-mail: larionov@fias.uni-frankfurt.de}

\affiliation{Frankfurt Institute for Advanced Studies (FIAS), 
             D-60438 Frankfurt am Main, Germany} 
\affiliation{National Research Centre "Kurchatov Institute", 
             123182 Moscow, Russia}

\begin{abstract}
Antiproton-induced reactions on nuclei at the beam energies from hundreds
MeV up to several GeV provide an excellent opportunity to study interactions 
of the antiproton and secondary particles (mesons, baryons and antibaryons) 
with nucleons. 
Antiproton projectile is unique in the sense that most of annihilation
particles are relatively slow in the target nucleus frame. Hence, 
the prehadronic effects do not much influence their interactions with the
nucleons of the nuclear residue.
Moreover, the particles with momenta less than about 1 GeV/c
are sensitive to the nuclear mean field potentials.
This paper discusses the microscopic transport calculations of the 
antiproton-nucleus reactions and is focused on three related problems:
(i) the antiproton potential determination, (ii) possible formation 
of strongly bound antiproton-nucleus systems, and (iii) strangeness production.
\end{abstract}


\maketitle

\section{Motivation}

It is difficult to produce antiproton beams. However, antiproton-nucleus interactions 
attract experimentalists and theorists since about 30 years when the KEK and LEAR data 
appeared. Since this time significant progress has been done to describe these data on the basis of optical
and cascade models. Still, antiproton interactions inside nuclei remain to be better understood. 
One example is the antiproton-nucleus optical potential. According to the low-density theorem,  
it can be expressed as
\begin{equation}
     V_{\rm opt} = -\frac{2\pi\sqrt{s}}{E_{\bar p} E_p} f_{\bar p p}(0) \rho~,    \label{Vopt}
\end{equation}
where at threshold $\sqrt{s} \simeq 2m_N$, $E_{\bar p} \simeq m_N$, $f_{\bar p p} \simeq (-0.9+i0.9)$ fm 
\cite{Batty:1997zp}. Being extrapolated to the normal nuclear density $\rho_0=0.16$ fm$^{-3}$, 
Eq.(\ref{Vopt}) predicts the repulsive antiproton-nucleus potential, $\mbox{Re} V_{\rm opt} \simeq 75$ MeV.  
In contrast, the $\bar p$-atomic X-ray and radiochemical data analysis \cite{Friedman:2005ad}
favors the strongly attractive antiproton-nucleus potential, $\mbox{Re} V_{\rm opt} \simeq -100$ MeV 
in the nuclear center.
Thus the $\bar p A$ optical potential is not a simple superposition of vacuum $\bar p N$ interactions.
The strongly attractive $\bar p A$ potential is consistent with Relativistic Mean Field (RMF) models
and has a consequence that a nucleus may collectively respond on the presence of an implanted antiproton.
The formation of strongly bound $\bar p$-nuclei becomes possible \cite{Buervenich:2002ns,Mishustin:2004xa}.

Another very interesting aspect is $\bar p$-annihilation in the nuclear interior. This results in a large
energy deposition $\geq 2 m_N$ in the form of mesons, mostly pions, in a volume of hadronic size $\sim 1-2$ fm
\cite{Rafelski:1979nt,Mishustin:2004xa}. After the passage of annihilation hadrons through the nuclear medium 
a highly excited nuclear residue can be formed and even 
experience explosive multifragment breakup \cite{Rafelski:1979nt,Cugnon:1986tx}.  
The annihilation of an antiproton at $p_{\rm lab} < \sim 5$ GeV/c 
on a nuclear target gives an excellent opportunity to study the interactions of secondary particles
(pions  \cite{Ilinov:1982jk},  kaons and hyperons  \cite{Ko:1987hf}, charmonia  \cite{Brodsky:1988xz,Farrar:1988me})
with nucleons. This is because most of annihilation hadrons are slow ($\gamma < 2$) and have short formation lengths.
Thus their interactions are governed by usual hadronic cross sections.   

Over last decades several microscopic transport models have been developed to describe
particle production in $\bar pA$ interactions \cite{Ilinov:1982jk,Cugnon:1986tx,Cugnon:1990xw,Strottman:1985jp,Gibbs:1990sf}.
Nowadays there is a renaissance in this field, since the antiproton-nucleus reactions 
at $p_{\rm lab} \simeq 1.5-15$ GeV/c will be a part of the  PANDA experiment at FAIR.
The most recent calculations are done within the Giessen Boltzmann-Uehling-Uhlenbeck (GiBUU) model
\cite{Larionov:2008wy,Larionov:2009tc,Larionov:2011fs} and within the Lanzhou quantum molecular dynamics (LQMD) model
\cite{Feng:2014cqa,Feng_IWND2014}. In the present paper I will report some results of GiBUU calculations 
for $\bar p$-nucleus interactions at $p_{\rm lab} \simeq 0.1-15$ GeV/c.

\section{GiBUU model}
\label{model}

The GiBUU model \cite{Buss:2011mx,GiBUU} solves a coupled set of kinetic equations for baryons, antibaryons,
and mesons.  In a RMF mode, this set can be written as (c.f. Refs. \cite{Ko:1987gp,Blaettel:1993uz})
\begin{eqnarray}
&& (p^{* 0})^{-1}
  \left[ p^{* \mu} \partial_\mu + ( p_{\mu}^* {\cal F}_j^{\alpha\mu}
    + m_j^* \partial^\alpha  m_j^* )
    \frac{\partial}{\partial p^{* \alpha}} \right]
 f_j(x,\bvec{p}^*)    \nonumber \\
&&= I_j[\{f\}] ~,      \label{BUUstar}
\end{eqnarray}
where $\alpha=1,2,3$, $\mu=0,1,2,3$, $x=(t,\mathbf{r})$;  
$j=N,\,\bar N,\,\Delta,\,\bar\Delta,\,Y,\bar Y,\,\pi,\,K,\,\bar K$ etc..
$f_j(x,\bvec{p}^*)$ is the distribution function of the particles of sort $j$ 
normalized such that the total number of particles of this sort is
\begin{equation}
    \int\, \frac{g_j d^3 r d^3 p^*}{(2\pi)^3} f_j(x,\bvec{p}^*)~,  \label{norma}
\end{equation}
with $g_j$ being the spin degeneracy factor. The Vlasov term (the l.h.s. of Eq.(\ref{BUUstar})) describes 
the evolution of the distribution function in smooth mean field potentials. The collision term
(the r.h.s. of Eq.(\ref{BUUstar})) accounts for elastic and inelastic binary collisions and  resonance 
decays. The Vlasov term includes the effective (Dirac) mass $m_j^*=m_j+S_j$, 
where $S_j=g_{\sigma j}\sigma$ is a scalar field;  
the field tensor ${\cal F}_j^{\mu\nu} = \partial^\mu V_j^\nu - \partial^\nu V_j^\mu$, 
where  $V_j^\mu = g_{\omega j} \omega^\mu + g_{\rho j} \tau^3 \rho^{3\mu} + q_j A^\mu$ is a vector field,
$\tau^3=+1$ for $p$ and $\bar n$, $\tau^3=-1$ for $\bar p$ and $n$; and the kinetic four-momentum 
$p^{* \mu}=p^\mu-V_j^\mu$  satisfying the effective mass shell condition $p^{* \mu}p^*_\mu={m_j^*}^2$.

In the present calculations, the nucleon-meson coupling constants $g_{\sigma N}, g_{\omega N}, g_{\rho N}$
and the self-interaction parameters of the $\sigma$-field have been adopted from a non-linear Walecka model
in the NL3 parameterization \cite{Lalazissis:1996rd}. The latter gives the compressibility 
coefficient $K=271.76$ MeV and the nucleon effective mass $m_N^*=0.60\, m_N$ 
at $\rho=\rho_0$. The antinucleon-meson coupling constants have been determined as
\begin{equation}
g_{\omega \bar N} = -\xi g_{\omega N},~g_{\rho \bar N} = \xi g_{\rho N},~
g_{\sigma \bar N} = \xi g_{\sigma N}~,                        \label{coupling_const}
\end{equation}
where $0 < \xi \leq 1$ is a scaling factor. The choice $\xi=1$ corresponds to the $G$-parity transformed
nuclear potential. In this case, however, the Schr\"odinger equivalent potential 
\begin{equation}
   U_{\bar N} = S_{\bar N} + V_{\bar N}^0 + \frac{(S_{\bar N})^2-(V_{\bar N}^0)^2}{2m_N}       \label{U_barN}
\end{equation}
becomes unphysically deep, $U_{\bar N} = -660$ MeV. The empirical choice of $\xi$ will be discussed
in the following section.

The GiBUU collision term includes the following channels\footnote{The GiBUU code is constantly developing. Thus the actual
version may include more channels. This description approximately corresponds to the release 1.4.0.}
(notations: $B$ -- nonstrange baryon, $R$ -- nonstrange baryon resonance, $Y$ -- hyperon with $S=-1$, 
$M$ -- nonstrange meson): 

\begin{itemize}

\item Baryon-baryon collisions:\\
elastic (EL) and charge-exchange (CEX) scattering $B B \to B B$; 
s-wave pion production/absorption\footnote{Implemented in a non-RMF mode only.} $NN \leftrightarrow NN\pi$; 
$NN \leftrightarrow \Delta\Delta$;  $NN \leftrightarrow NR$; $N(\Delta,N^*) N(\Delta,N^*) \to N(\Delta)YK$;
$Y N \to Y N$; $\Xi N \to \Lambda \Lambda$; $\Xi N \to \Lambda \Sigma$;
$\Xi N \to \Xi N$.\\
For invariant energies $\sqrt{s} > 2.6$ GeV the inelastic production 
$B_1 B_2 \to B_3 B_4$ (+ mesons) is  simulated via the PYTHIA model.

\item Antibaryon-baryon collisions:\\
annihilation $\bar B B \to$ mesons\footnote{Described with a help of the statistical annihilation model 
\cite{Golubeva:1992tr,PshenichnovPhD}.};  EL and CEX scattering  $\bar B B \to \bar B B$;
$\bar N N \leftrightarrow \bar N \Delta$ (+ c.c.);
$\bar N N \to \bar\Lambda \Lambda$; 
$\bar N(\bar\Delta) N(\Delta) \to \bar\Lambda \Sigma$ (+ c.c.);
$\bar N(\bar\Delta) N(\Delta) \to \bar\Xi \Xi$.\\
For invariant energies $\sqrt{s} > 2.4$ GeV (i.e. $p_{\rm lab} > 1.9$ GeV/c for $\bar N N$)
the inelastic production $\bar B_1 B_2 \to \bar B_3 B_4$ (+ mesons) is simulated via the
FRITIOF model.

\item Meson-baryon collisions:\\
$M N \leftrightarrow R$ (baryon resonance excitations and decays, e.g. $\pi N \leftrightarrow \Delta$
and $\bar K N \leftrightarrow Y^*$); 
$\pi(\rho) \Delta \leftrightarrow R$;  
$\pi N \to \pi N$; $\pi N \to \pi\pi N$; $\pi N \to \eta \Delta$;
$\pi N \to \omega N$; $\pi N \to \phi N$; 
$\pi N \to \omega \pi N$; $\pi N \to \phi \pi N$;
$\pi(\eta,\rho,\omega) N \to YK$;  $\pi N \to K \bar K N$; 
$\pi N \to Y K \pi$; $\pi \Delta \to Y K$;
$K N \to K N$ (EL, CEX);
$\bar K N \to \bar K N$ (EL, CEX);
$\bar K N \leftrightarrow Y \pi$; $\bar K N \leftrightarrow Y^* \pi$;
$\bar K N \to \Xi K$.\\
At $\sqrt{s} > 2.2$ GeV the inelastic meson-baryon collisions are simulated 
via  PYTHIA.

\item Meson-meson collisions:\\
$M_1 M_2 \leftrightarrow M_3$ (meson resonance excitations and decays, 
e.g. $\pi \pi \leftrightarrow \rho$ and $K \pi \leftrightarrow K^*$);
$M_1 M_2 \leftrightarrow K\overline K$, $M_1 M_2 \leftrightarrow K \overline K^*$ (+ c.c.).

\end{itemize}

\section{Antiproton absorption and annihilation on nuclei}
\label{absorption}

\begin{figure}
   \includegraphics[width=\columnwidth]{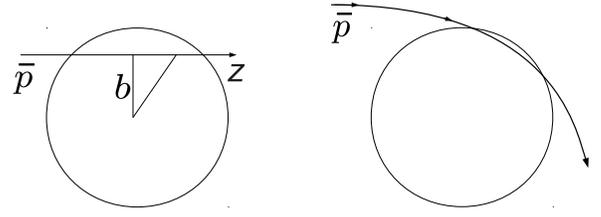}
\caption{\label{fig:curved} Left panel -- the straight-line propagation of an antiproton 
in the absence of a mean field. Right panel -- the illustration of the curved trajectory of an 
antiproton due to an attractive mean field.}
\end{figure}

Without mean field acting on an antiproton the GiBUU model is expected to reproduce a simple Glauber 
model result for the $\bar p$-absorption cross section on a nucleus (left Fig.~\ref{fig:curved}):
\begin{equation}
   \sigma_{\rm abs}^{\rm Glauber}=\int d^2b
        \left(1 - \mbox{e}^{-\overline\sigma_{\rm tot} \int\limits_{-\infty}^{+\infty}\,dz\rho(b,z)}\right)~,
                                          \label{sigabs^Glauber}
\end{equation}
where $\overline\sigma_{\rm tot}$ is the isospin-averaged total $\bar p N$ cross section. 
The attractive mean field bends the $\bar p$ trajectory to the nucleus (right Fig.~\ref{fig:curved}).
Thus the absorption cross section should increase.

\begin{figure}
   \includegraphics[width=\columnwidth]{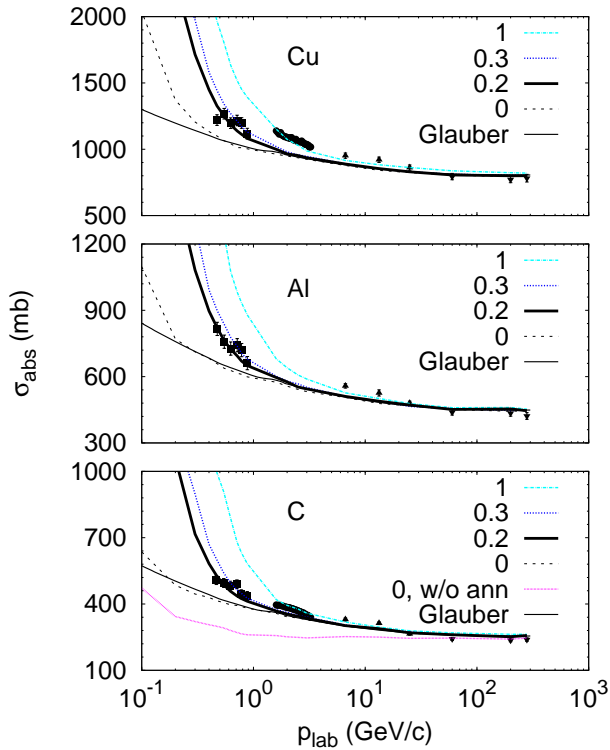}
\caption{\label{fig:sigabs} Antiproton absorption cross section on  
the $^{12}$C, $^{27}$Al, and $^{64}$Cu nuclei vs the beam momentum. The GiBUU results are shown by 
the lines marked with the value of a scaling factor $\xi$.
Thin solid lines represent the Glauber model calculation, Eq.(\ref{sigabs^Glauber}). 
For the $\bar p$+$^{12}$C system, a calculation with $\xi=0$ without annihilation 
is shown by the dotted line.}
\end{figure}

Fig.~\ref{fig:sigabs} shows the GiBUU calculations of antiproton absorption cross sections
on $^{12}$C, $^{27}$Al and $^{64}$Cu in comparison with experimental data 
\cite{Nakamura:1984xw,Abrams:1972ab,Denisov:1973zv,Carroll:1978hc} and with
the Glauber formula (\ref{sigabs^Glauber}).
Indeed, GiBUU calculations without mesonic components of the $\bar p$ mean field, i.e.
with scaling factor $\xi=0$, are very close to Eq.(\ref{sigabs^Glauber}) 
at $p_{\rm lab} > 0.3$ GeV/c. At lower $p_{\rm lab}$, the Coulomb potential makes
the difference between GiBUU ($\xi=0$) and Glauber results. Including the mesonic components 
of $\bar p$ mean field ($\xi > 0$) noticeably increases the absorption cross section 
at $p_{\rm lab} < 3$ GeV/c.
The best fit of the KEK data \cite{Nakamura:1984xw} at $p_{\rm lab}=470-880$ MeV/c is reached with
$\xi=0.21\pm0.03$. This produces the real part of the antiproton-nucleus optical potential
$\mbox{Re} V_{\rm opt} \equiv U_{\bar p} \simeq -(150\pm30)$ MeV at $\rho=\rho_0$.
The corresponding imaginary part is
\begin{equation}
     \mbox{Im} V_{\rm opt} = -\frac{1}{2} <v_{\bar p N} \overline\sigma_{\rm tot}> \rho~.  \label{ImVopt}
\end{equation}
At  $\rho=\rho_0$ this gives $\mbox{Im} V_{\rm opt} \simeq -(100-110)$ MeV independent 
on the choice of $\xi$. It is interesting
that the BNL \cite{Abrams:1972ab} and Serpukhov \cite{Denisov:1973zv} data at 
$p_{\rm lab}=1.6-20$ GeV/c favor $\xi=1$, i.e.  $\mbox{Re} V_{\rm opt} \simeq -660$ MeV
at $\rho=\rho_0$. This discrepancy needs to be clarified which could be possibly done at FAIR.

\begin{figure}
   \includegraphics[width=\columnwidth]{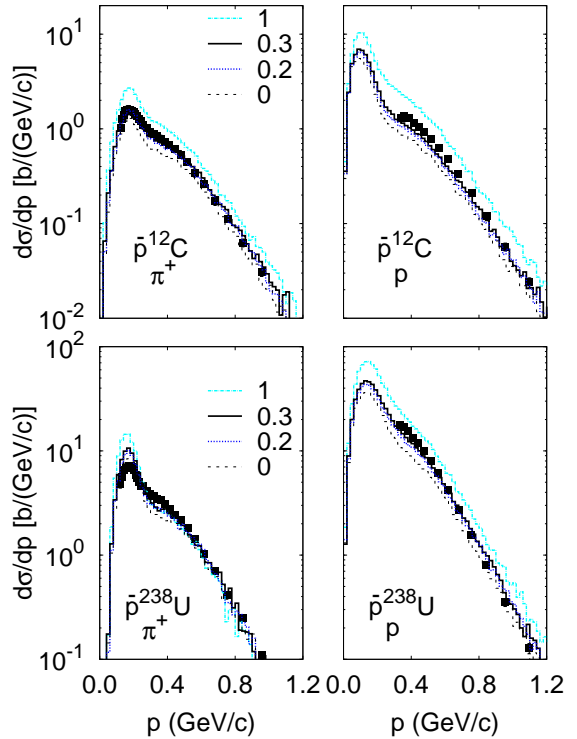}
\caption{\label{fig:180mev_mom} Momentum differential cross sections of $\pi^+$ and $p$
production in $\bar p$ annihilation at 608 MeV/c on $^{12}$C and $^{238}$U. The different
lines are denoted by the value of a scaling factor $\xi$. The data points are from 
\cite{Mcgaughey:1986kz}.}
\end{figure}

Fig.~\ref{fig:180mev_mom} displays the calculated momentum spectra of positive pions and protons
for antiproton interactions at $p_{\rm lab}=608$ MeV/c with the carbon and uranium targets. 
GiBUU very well reproduces a quite complicated shape of the
pion spectra which appears due to the underlying $\pi N \leftrightarrow \Delta$ dynamics.
The absolute normalization of the spectra is weakly sensitive to the $\bar p$ mean field.
The best agreement is reached for $\xi=0.3$, i.e. for $\mbox{Re} V_{\rm opt} \simeq -(220\pm70)$ MeV.

\section{Selfconsistency effects}
\label{selfconst}

Strong attraction of an antiproton to the nucleus has to influence on the nucleus itself.
This back coupling effect can be taken into account by including the
antinucleon contributions in the source terms of the Lagrange equations for $\omega$-, $\rho$-,
and $\sigma$-fields:
\begin{eqnarray}
   & &  (\partial_\mu\partial^\mu + m_\omega^2)\omega^\nu(x)
                               =  \sum_{j=N,\bar N} g_{\omega j} \langle\bar\psi_j(x) \gamma^\nu \psi_j(x)\rangle,  \label{KGeqOmega} \\
   & &  (\partial_\mu\partial^\mu + m_\rho^2)\rho^{3\,\nu}(x) 
                               =  \sum_{j=N,\bar N} g_{\rho j} \langle\bar\psi_j(x) \gamma^\nu \tau^3 \psi_j(x)\rangle,  \label{KGeqRho} \\
   & &  \partial_\mu\partial^\mu \sigma(x) + \frac{dU(\sigma)}{d\sigma}
                               = - \sum_{j=N,\bar N} g_{\sigma j} \langle\bar\psi_j(x) \psi_j(x)\rangle,   \label{KGeqSigma}
\end{eqnarray}
with $U(\sigma)=\frac{1}{2}m_\sigma^2\sigma^2+\frac{1}{3}g_2\sigma^3+\frac{1}{4}g_3\sigma^4$,
or, in other words, by treating the meson fields selfconsistently. As follows from Eqs. (\ref{coupling_const}) and
(\ref{KGeqOmega})-(\ref{KGeqSigma}), nucleons and antinucleons contribute with the opposite sign to the source terms of the vector 
fields $\omega$ and $\rho$, and with the same sign -- to the source term of the scalar field $\sigma$.
Hence, repulsion is reduced and attraction is enhanced in the presence of an antiproton in the nucleus.

\begin{figure}
   \includegraphics[width=\columnwidth]{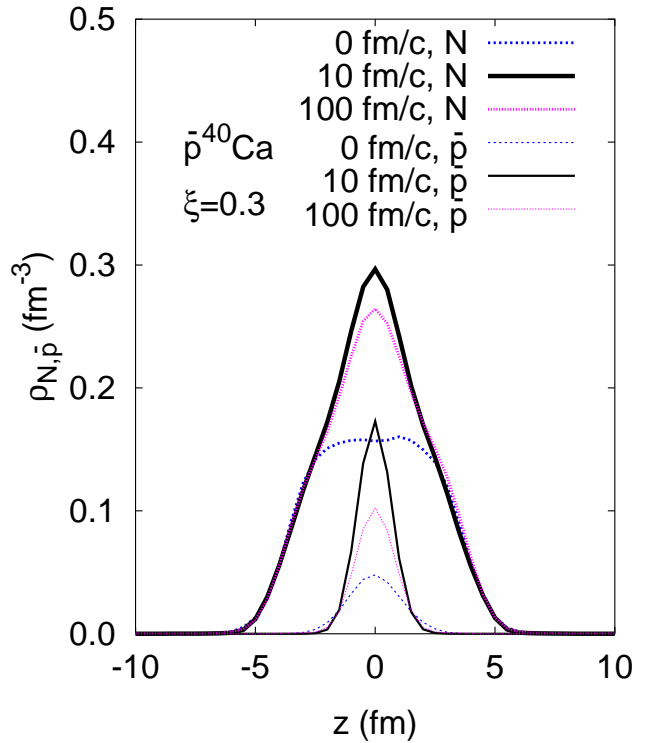}
\caption{\label{fig:rhoz_xi03_Ca40} The density of nucleons (thick lines) and antiproton (thin lines) as a function of coordinate on
$z$-axis drawn through the nuclear center ($z=0$).}
\end{figure}

Fig.~\ref{fig:rhoz_xi03_Ca40} shows the density profiles of nucleons and of an antiproton at the different time moments for the case
of the $\bar p$ implanted at $t=0$ in the center of the $^{40}$Ca nucleus. As the consequence of a pure Vlasov dynamics of the coupled
antiproton-nucleus system (annihilation is turned off), both the nucleon and the antiproton densities grow quite fast.
At $t \sim 10$ fm/c the compressed state is already formed, and the system starts to oscillate around the new equilibrium density
$\rho \simeq 2\rho_0$.

\begin{figure}
   \includegraphics[width=\columnwidth]{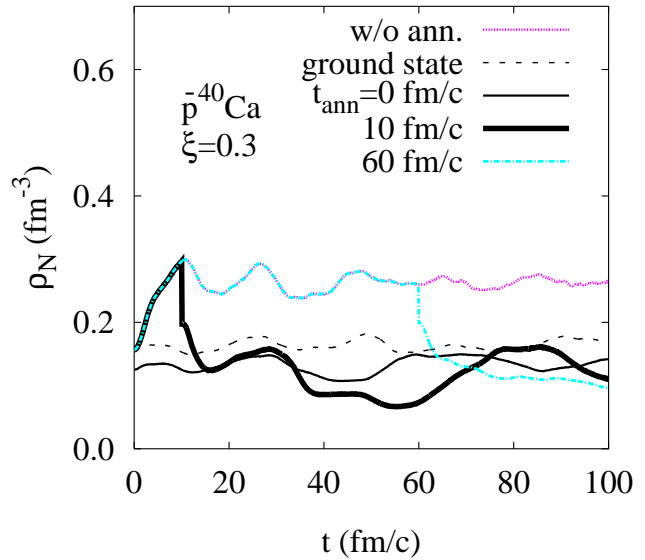}
\caption{\label{fig:rhoc_xi03_Ca40} The central nucleon density as a function of time. The annihilation of $\bar p$ with the closest nucleon into mesons 
is simulated at the time moment $t_{\rm ann}$ as indicated. The calculations without annihilation and for the ground state nucleus 
(without $\bar p$) are also shown.}
\end{figure}

Fig.~\ref{fig:rhoc_xi03_Ca40} displays the time evolution of the central nucleon density. The $\bar p$ annihilation is simulated at the time moment $t_{\rm ann}$.
The choice  $t_{\rm ann}=0$ corresponds to the usual annihilation of a stopped $\bar p$ in the nuclear center. In this case, the nucleon
density remains close to the ground state density. However, if the annihilation is simulated in a compressed configuration ($t_{\rm ann} > 0$),
then the residual nuclear system expands. Eventually the system  reaches the low-density spinodal region ($\rho < \sim 0.6\rho_0$), where the
sound velocity squared $c_s^2=\partial P/\partial \rho_{|s=\mbox{const}}$ becomes negative\footnote{Here, $P$ is the pressure and $s$ is the entropy per nucleon.}.
This should result in the breakup of the residual nuclear system into fragments.

\begin{figure}
   \includegraphics[width=\columnwidth]{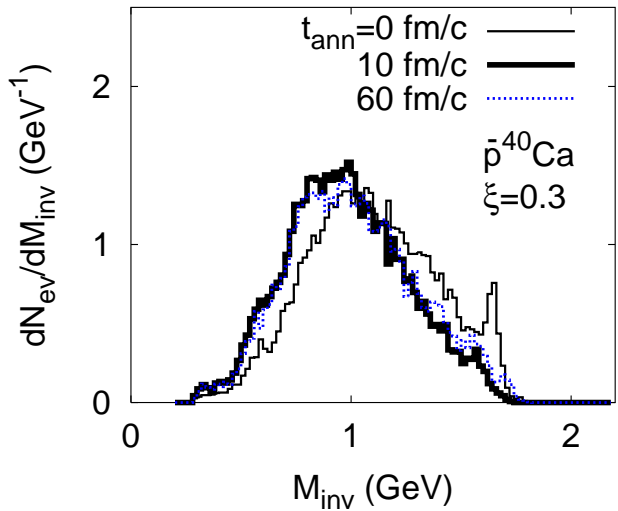}
\caption{\label{fig:dmdiMes_xi03_Ca40} Annihilation event spectrum on the total invariant mass (\ref{Minv}) of emitted
mesons. Calculations are done for three different values of annihilation time $t_{\rm ann}$.}
\end{figure}

A possible observable signal of the $\bar p$ annihilation in a compressed nuclear configuration is the total invariant
mass $M_{\rm inv}$ of emitted mesons 
\begin{equation}
     M_{\rm inv}^2 = \left( \sum\limits_i p_i \right)^2~.             \label{Minv}
\end{equation}         
For the annihilation of a stopped antiproton on a proton at rest in vacuum, $M_{\rm inv}=2m_N$. 
In nuclear medium, the proton and antiproton vector fields largely cancel each other 
\footnote{The cancellation is exact for the antiproton vector fields obtained
by the $G$-parity transformation from the respective proton vector fields, i.e. when $\xi=1$.}. 
Therefore, it is expected that in nuclear medium the peak will appear at $M_{\rm inv} \simeq 2m_N^*$. 
This simple picture is illustrated by GiBUU calculations in Fig.~\ref{fig:dmdiMes_xi03_Ca40}.
In calculations with $t_{\rm ann}=0$ we clearly see a sharp medium-modified peak shifted downwards by $\simeq 200$ MeV
from $2m_N$. The final state interactions of mesons make a broad maximum at $M_{\rm inv} \simeq 1$ GeV.
For annihilation in compressed configurations ($t_{\rm ann}=10$ and 60 fm/c), the total spectrum further shifts by about $100$ MeV
to smaller $M_{\rm inv}$. This effect becomes stronger with decreasing mass of the target nucleus (e.g., for $^{16}$O 
the spectrum shift is nearly $500$ MeV \cite{Larionov:2008wy}).    

\section{Strangeness production}
\label{strangeness}

Originally, the main motivation of experiments on strangeness production in antiproton-nucleus collisions was to find 
the signs of unusual phenomena, in-particular, of a multinucleon annihilation and/or of a quark-gluon plasma (QGP) formation. 
In Ref. \cite{Rafelski:1988wn}, the cold QGP formation has been suggested to explain the unusually large ratio 
$\Lambda/K^0_S \simeq 2.4$ measured in the reaction $\bar p^{181}$Ta at 4 GeV/c \cite{Miyano:1984dc}.
On the other hand, in Refs. \cite{Ko:1987hf,Cugnon:1990xw,Ahmad:1997fv,Larionov:2011fs,Gaitanos:2011fy,%
Feng:2014cqa,Feng_IWND2014,Gaitanos:2014bla} 
most features of strangeness production in $\bar pA$ reactions have been explained by hadronic mechanisms.

\begin{figure}
   \includegraphics[width=\columnwidth]{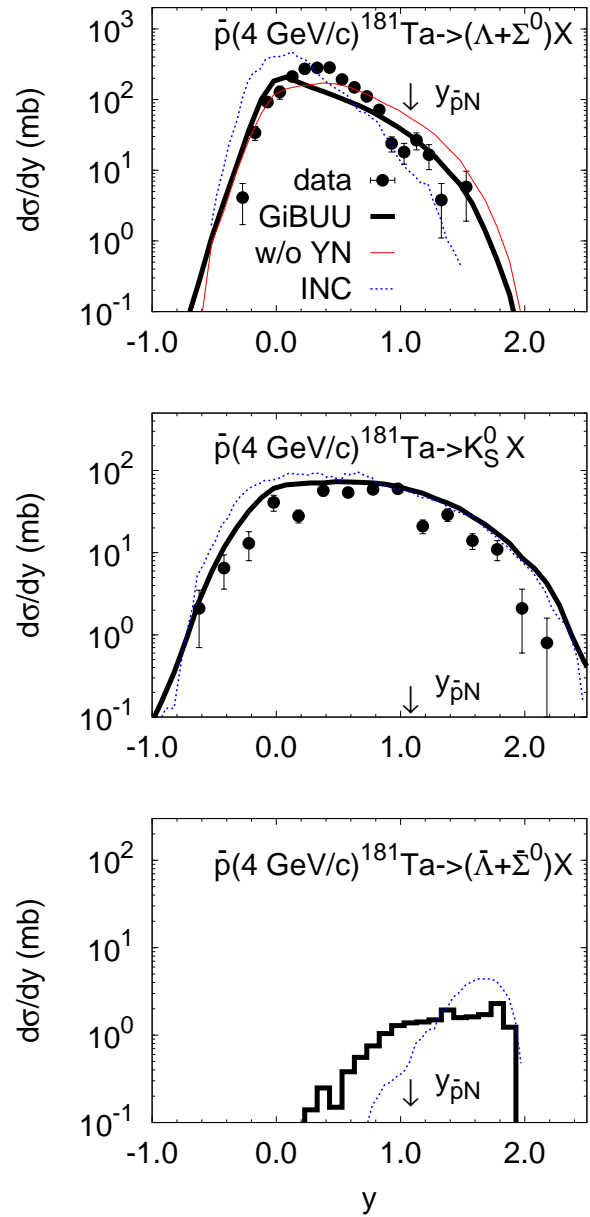}
\caption{\label{fig:dsig_dy_pbarTa} Rapidity spectra of $(\Lambda+\Sigma^0)$, $K^0_S$, and  $(\bar\Lambda+\bar\Sigma^0)$
from  $\bar p^{181}$Ta collisions at 4 GeV/c. See text for details.}
\end{figure}

Fig.~\ref{fig:dsig_dy_pbarTa} presents the rapidity spectrum of $(\Lambda+\Sigma^0)$ hyperons, $K^0_S$ mesons and  
$(\bar\Lambda+\bar\Sigma^0)$ antihyperons for collisions $\bar p(\mbox{4 GeV/c})^{181}$Ta in comparison
with the data \cite{Miyano:1988mq} and the intranuclear cascade (INC) calculations \cite{Cugnon:1990xw}.
The GiBUU model underpredicts hyperon yields at small forward rapidities $y \simeq 0.5$ and overpredicts
$K^0_S$ yields. In the GiBUU calculation without hyperon-nucleon scattering, the $(\Lambda+\Sigma^0)$
spectrum is shifted to forward rapidities. However, the problem of underpredicted total $(\Lambda+\Sigma^0)$ yield
remains. A more detailed analysis \cite{Larionov:2011fs} shows that 72\% of $Y$ and $Y^*$ production
rate in GiBUU is due to antikaon absorption processes $\bar K B \to YX$, $\bar K B \to Y^*$, and $\bar K B \to Y^* \pi$.
The second largest contribution, $23\%$ of the rate, is caused by the nonstrange meson - baryon collisions.
The antibaryon-baryon (including the direct $\bar p N$ channel) and baryon-baryon collisions contribute only
3\% and 2\%, respectively, to the same rate. 
The underprediction of the hyperon yield in GiBUU could be due to the used partial $\bar K N$ cross sections,
in-particular, due to the problematic $K^- n$ channel\footnote{The $K^- n$ channel has been improved in recent
GiBUU releases, however, after the present calculations were already done.}.
The possible in-medium enhancement of the hyperon production in antikaon-baryon collisions is also not excluded.

\begin{figure}
   \includegraphics[width=\columnwidth]{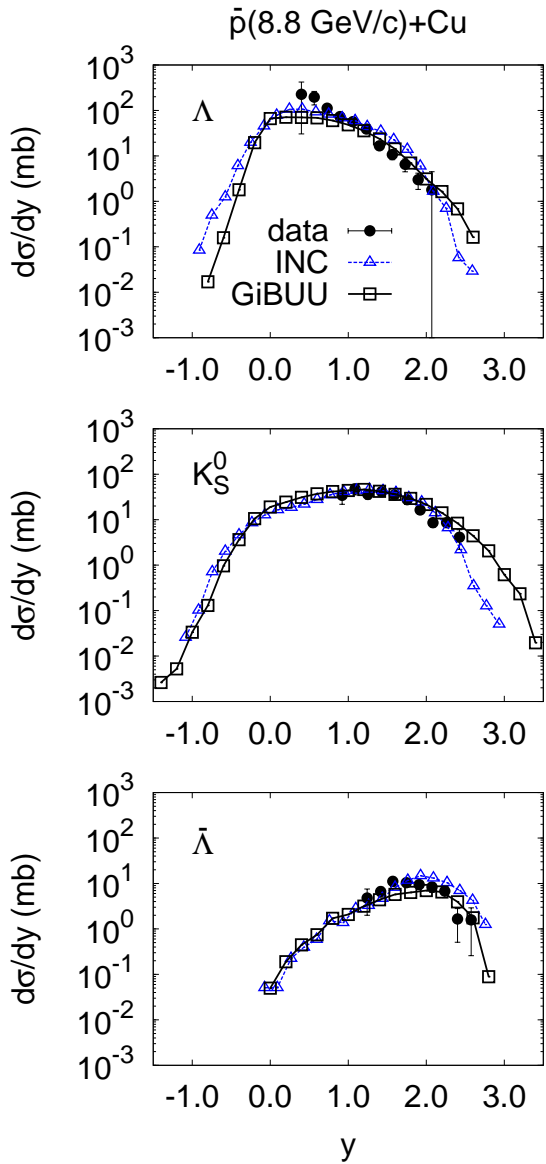}
\caption{\label{fig:dsig_dy_pbarCu} Rapidity spectra of $(\Lambda+\Sigma^0)$, $K^0_S$, and  $(\bar\Lambda+\bar\Sigma^0)$
from $\bar p^{64}$Cu collisions at 8.8 GeV/c. The data and INC calculations are from \cite{Ahmad:1997fv}.}
\end{figure}

As shown in Fig.~\ref{fig:dsig_dy_pbarCu}, at higher beam momenta the agreement between the calculations 
and the data on neutral strange particle production becomes visibly better. Exception is again the region 
of small forward rapidities $y \simeq 0.5$ where both GiBUU and INC calculations underpredict the $(\Lambda+\Sigma^0)$
yield.

\begin{figure}
   \includegraphics[width=\columnwidth]{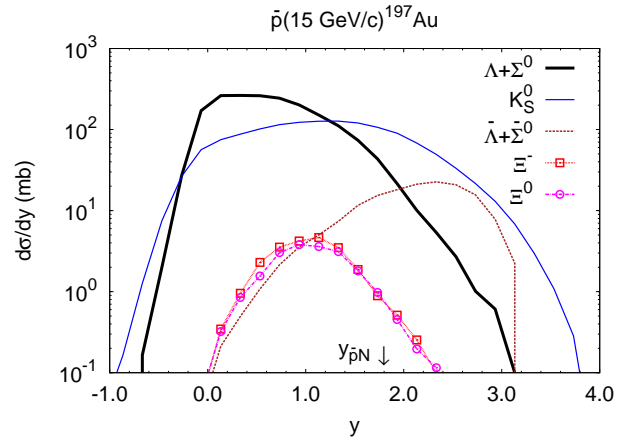}
\caption{\label{fig:dsig_dy_pbarAu} The rapidity spectra of $(\Lambda+\Sigma^0)$, $K^0_S$, $(\bar\Lambda+\bar\Sigma^0)$,
$\Xi^-$, and $\Xi^0$ from $\bar p^{197}$Au collisions at 15 GeV/c.}
\end{figure}

Finally, let us discuss the $\Xi$ ($S=-2$) hyperon production. The direct production of $\Xi$ in the collision of nonstrange
particles would require to produce two $s \bar s$ pairs simultaneously. Thus, $\Xi$ production could be even stronger enhanced
in a QGP as compared to the enhancement for the $S=-1$ hyperons. 
Fig.~\ref{fig:dsig_dy_pbarAu} shows the rapidity spectra of the different strange particles in $\bar p ^{197}$Au collisions at 15 GeV/c.
Even at such a high beam momentum, the $S=-1$ hyperon spectra still have a flat maximum at $y \simeq 0$ due to {\it exothermic}
strangeness exchange reactions $\bar K N \to Y \pi$ with slow $\bar K$. In contrast, the second largest, $\sim 18\%$, contribution
to the $\Xi$ production is given by  {\it endothermic} double strangeness exchange reactions $\bar K N \to \Xi K$
\footnote{The main, $\sim 24\%$, contribution to the total yield of $\Xi$'s at 15 GeV/c is given by $\Xi^* \to \Xi \pi$ decays.
The direct channel $\bar N N \to \bar\Xi \Xi$ contributes $\sim 10\%$ only.}. Since the threshold beam momentum of $\bar K$
for the process $\bar K N \to \Xi K$ is $1.05$ GeV/c, which corresponds to the $\bar KN$ c.m. rapidity of 0.55, 
the rapidity spectra of $\Xi$'s are shifted forward with respect to the $\Lambda$ rapidity spectra. However, in the QGP
fireball scenario \cite{Rafelski:1988wn}, the rapidity spectra of all strange particles would be peaked at the same rapidity.

\section{Summary}
\label{summary}

This work was focused on the dynamics of a coupled antiproton-nucleus system and on the strangeness production 
in $\bar pA$ interactions. The calculations were based on the GiBUU transport model.
The main results can be summarized as:
\begin{itemize}

\item The reproduction of experimental data on $\bar pA$ absorption cross sections at $p_{\rm lab} < 1$ GeV/c
and on $\pi^+$ and $p$ production at $p_{\rm lab}=608$ MeV/c requires to use a strongly attractive
$\bar pA$ optical potential, $\mbox{Re} V_{\rm opt} \simeq -(150-200)$ MeV at $\rho=\rho_0$.

\item As the response of a nucleus to the presence of an antiproton, the nucleon density can be
increased up to $\rho \sim (2-3)\rho_0$  locally near $\bar p$. Annihilation of the $\bar p$ in such a compressed
configuration can manifest itself in the multifragment breakup of the residual nuclear system and in the
substantial ($\sim 300-500$ MeV) shift of annihilation event spectrum on the total invariant mass of
produced mesons $M_{\rm inv}$ toward low $M_{\rm inv}$.

\item GiBUU describes the data on inclusive pion and proton production fairly well. Still, the strangeness
production remains to be better understood (overestimated $K^0_S$ - and underestimated $(\Lambda+\Sigma^0)$ -
production).

\item $\Xi$ hyperon forward rapidity shift with respect to $\Lambda$ is suggested as a test of
hadronic and QGP mechanisms of strangeness production in $\bar pA$ reactions.

\end{itemize}

\begin{acknowledgments}
The author is grateful to T. Gaitanos,  W. Greiner, \mbox{I.~N.~Mishustin}, U. Mosel, I.~A.~Pshenichnov, 
and L.~M.~Satarov for the collaboration on the GiBUU studies of $\bar pA$ reactions.
This work was supported by HIC for FAIR within the framework of the LOEWE program.
\end{acknowledgments}


\begin{thebibliography}{99}

\bibitem{Batty:1997zp} Batty C J, Friedman E, Gal A. Strong interaction physics from hadronic atoms.
Phys Rep, 1997, {\bf 287}: 385--445. \href{http://dx.doi.org/10.1016/S0370-1573(97)00011-2}{DOI: 10.1016/S0370-1573(97)00011-2}

\bibitem{Friedman:2005ad} Friedman E, Gal A, Mare\v{s} J. Antiproton nucleus potentials from global fits to
antiprotonic X-rays and radiochemical data. Nucl Phys A, 2005, {\bf 761}: 283--295.
\href{http://dx.doi.org/10.1016/j.nuclphysa.2005.08.001}{DOI: 10.1016/j.nuclphysa.2005.08.001}

\bibitem{Buervenich:2002ns} B\"urvenich T, Mishustin I N, Satarov L M, \emph{et al}. 
Enhanced binding and cold compression of nuclei due to admixture of anti-baryons.
Phys Lett B, 2002, {\bf 542}: 261--267.
\href{http://dx.doi.org/10.1016/S0370-2693(02)02351-1}{DOI: 10.1016/S0370-2693(02)02351-1}

\bibitem{Mishustin:2004xa} Mishustin I N, Satarov L M, B\"urvenich T J, St\"ocker H, Greiner W.
Antibaryons bound in nuclei. Phys Rev C, 2005, {\bf 71}:  035201.
\href{http://dx.doi.org/10.1103/PhysRevC.71.035201}{DOI: 10.1103/PhysRevC.71.035201}

\bibitem{Rafelski:1979nt} Rafelski J. $\bar p$ annihilation on heavy nuclei.
Phys Lett B, 1980, {\bf 91}: 281--284.
\href{http://dx.doi.org/10.1016/0370-2693(80)90450-5}{10.1016/0370-2693(80)90450-5}

\bibitem{Cugnon:1986tx} Cugnon J, Vandermeulen J. 
Transfer of energy following $\bar p$-annihilation on nuclei.
Nucl Phys A, 1985, {\bf 445}: 717--736.
\href{http://dx.doi.org/10.1016/0375-9474(85)90568-8}{10.1016/0375-9474(85)90568-8}

\bibitem{Ilinov:1982jk} Iljinov A S, Nazaruk V I, Chigrinov S E.
Nuclear absorption of stopped antiprotons: Multipion-nucleus interactions.
Nucl Phys A, 1982, {\bf 382}: 378--400.
\href{http://dx.doi.org/10.1016/0375-9474(82)90352-9}{10.1016/0375-9474(82)90352-9}

\bibitem{Ko:1987hf} Ko C M, Yuan R. Lambda production from anti-proton annihilation in nuclei.
Phys Lett B, 1987, {\bf 192}: 31--34.
\href{http://dx.doi.org/10.1016/0370-2693(87)91136-1}{10.1016/0370-2693(87)91136-1}

\bibitem{Brodsky:1988xz} Brodsky S J, Mueller A H.
Using nuclei to probe hadronization in QCD.
Phys. Lett. B, 1988, {\bf 206}: 685-–690.
\href{http://dx.doi.org/10.1016/0370-2693(88)90719-8}{DOI: 10.1016/0370-2693(88)90719-8}

\bibitem{Farrar:1988me} Farrar G R, Liu H, Frankfurt L L, Strikman M I.
Transparency in Nuclear Quasiexclusive Processes with Large Momentum Transfer.
Phys Rev Lett, 1988, {\bf 61}: 686--689.
\href{http://dx.doi.org/10.1103/PhysRevLett.61.686}{DOI: 10.1103/PhysRevLett.61.686}

\bibitem{Cugnon:1990xw} Cugnon J, Deneye P, Vandermeulen J.
Strangeness production in antiproton annihilation on nuclei.
Phys Rev C, 1990, {\bf 41}: 1701--1718.
\href{http://dx.doi.org/10.1103/PhysRevC.41.1701}{DOI: 10.1103/PhysRevC.41.1701}

\bibitem{Strottman:1985jp} Strottman D, Gibbs W R.
High nuclear temperatures by antimatter-matter annihilation.
Phys Lett B, 1984, {\bf 149}: 288--292.
\href{http://dx.doi.org/10.1016/0370-2693(84)90408-8}{DOI: 10.1016/0370-2693(84)90408-8}

\bibitem{Gibbs:1990sf} Gibbs W R, Kruk J W.
Strangeness production in antiproton-tantalum interactions at 4 GeV/c.
Phys Lett B, 1990, {\bf 237}: 317--322.
\href{http://dx.doi.org/10.1016/0370-2693(90)91181-A}{DOI: 10.1016/0370-2693(90)91181-A}

\bibitem{Larionov:2008wy} Larionov A B, Mishustin I N, Satarov L M, Greiner, W.
Dynamical simulation of bound antiproton-nuclear systems and observable signals 
of cold nuclear compression. Phys Rev C, 2008, {\bf 78}: 014604.
\href{http://dx.doi.org/10.1103/PhysRevC.78.014604}{DOI: 10.1103/PhysRevC.78.014604}

\bibitem{Larionov:2009tc} Larionov A B, Pshenichnov I A, Mishustin I N, Greiner W.
Antiproton-nucleus collisions simulation within a kinetic approach with relativistic mean fields.
Phys Rev C, 2009, {\bf 80}: 021601.
\href{http://dx.doi.org/10.1103/PhysRevC.80.021601}{DOI: 10.1103/PhysRevC.80.021601}

\bibitem{Larionov:2011fs} Larionov A B, Gaitanos T, Mosel U.
Kaon and hyperon production in antiproton-induced reactions on nuclei.
Phys Rev C, 2012, {\bf 85}: 024614.
\href{http://dx.doi.org/10.1103/PhysRevC.85.024614}{DOI: 10.1103/PhysRevC.85.024614}

\bibitem{Feng:2014cqa} Feng Z Q, Lenske H.
Particle production in antiproton-induced nuclear reactions.
Phys Rev C, 2014, {\bf 89}: 044617.
\href{http://dx.doi.org/10.1103/PhysRevC.89.044617}{DOI: 10.1103/PhysRevC.89.044617}

\bibitem{Feng_IWND2014} Feng Z Q.
Nuclear dynamics induced by antiprotons.
Nucl Sci Tech, 2015, {\bf 26}: S20512.
\href{http://dx.doi.org/10.13538/j.1001-8042/nst.26.S20512}{DOI: 10.13538/j.1001-8042/nst.26.S20512}


\bibitem{Buss:2011mx} Buss O, Gaitanos T, Gallmeister K, \emph{et al}.
Transport-theoretical description of nuclear reactions.
Phys Rep, 2012, {\bf 512}: 1--124.
\href{http://dx.doi.org/10.1016/j.physrep.2011.12.001}{DOI: 10.1016/j.physrep.2011.12.001}

\bibitem{GiBUU} \href{https://gibuu.hepforge.org/trac/wiki}{https://gibuu.hepforge.org/trac/wiki}

\bibitem{Ko:1987gp} Ko C M, Li Q, Wang R C.
Relativistic Vlasov Equation for Heavy Ion Collisions.
Phys Rev Lett, 1987, {\bf 59}: 1084--1087.
\href{http://dx.doi.org/10.1103/PhysRevLett.59.1084}{DOI: 10.1103/PhysRevLett.59.1084}

\bibitem{Blaettel:1993uz} Bl\"attel B, Koch V, Mosel U.
Transport-theoretical analysis of relativistic heavy-ion collisions.
Rep Prog Phys, 1993, {\bf 56}: 1--62.
\href{http://dx.doi.org/10.1088/0034-4885/56/1/001}{DOI: 10.1088/0034-4885/56/1/001}

\bibitem{Lalazissis:1996rd} Lalazissis G A, K\"onig J, Ring P.
New parametrization for the Lagrangian density of relativistic mean field theory.
Phys Rev C, 1997, {\bf 55}: 540--543.
\href{http://dx.doi.org/10.1103/PhysRevC.55.540}{DOI: 10.1103/PhysRevC.55.540}

\bibitem{Golubeva:1992tr} Golubeva E S, Iljinov A S, Krippa B V, Pshenichnov I A.
Effects of mesonic resonance production in annihilation of stopped antiprotons on nuclei.
Nucl Phys A, 1992, {\bf 537}: 393--417.
\href{http://dx.doi.org/10.1016/0375-9474(92)90362-N}{DOI: 10.1016/0375-9474(92)90362-N}

\bibitem{PshenichnovPhD} Pshenichnov I A.
PhD thesis, INR, Moscow, 1998.

\bibitem{Nakamura:1984xw} Nakamura K, Chiba J, Fujii T, \emph{et al}.
Absorption and Forward Scattering of Antiprotons by C, Al, and Cu Nuclei in the Region 470-880 MeV/c.
Phys Rev Lett, 1984, {\bf 52}: 731--734.
\href{http://dx.doi.org/10.1103/PhysRevLett.52.731}{DOI: 10.1103/PhysRevLett.52.731}

\bibitem{Abrams:1972ab} Abrams R J, Cool R L, Giacomelli G, \emph{et al}.
Absorption Cross Sections of $K^\pm$ and $\bar p$ on Carbon and Copper in the Region 1.0-3.3 GeV/c.
Phys Rev D, 1971, {\bf 4}: 3235--3244.
\href{http://dx.doi.org/10.1103/PhysRevD.4.3235}{DOI: 10.1103/PhysRevD.4.3235}

\bibitem{Denisov:1973zv} Denisov S P, Donskov S V, Gorin Yu P, \emph{et al}.
Absorption cross sections for pions, kaons, protons and antiprotons on complex nuclei in the 6 to 60 GeV/c momentum range.
Nucl Phys B, 1973, {\bf 61}: 62--76.
\href{http://dx.doi.org/10.1016/0550-3213(73)90351-9}{DOI: 10.1016/0550-3213(73)90351-9}

\bibitem{Carroll:1978hc} Carroll A S, Chiang I H, Kycia T F, \emph{et al}.
Absorption cross section of $\pi^\pm$, $K^\pm$, $p$ and $\bar p$ on nuclei between 60 and 280 GeV/c.
Phys Lett B, 1979, {\bf 80}: 319--322.
\href{http://dx.doi.org/10.1016/0370-2693(79)90226-0}{DOI: 10.1016/0370-2693(79)90226-0}

\bibitem{Mcgaughey:1986kz} Mcgaughey P L, Bol K D, Clover M R, \emph{et al}.
Dynamics of Low-Energy Antiproton Annihilation in Nuclei as Inferred from 
Inclusive Proton and Pion Measurements.
Phys Rev Lett, 1986, {\bf 56}: 2156--2159.
\href{http://dx.doi.org/10.1103/PhysRevLett.56.2156}{DOI: 10.1103/PhysRevLett.56.2156}

\bibitem{Rafelski:1988wn} Rafelski J. 
Quark-gluon plasma in 4 GeV/c antiproton annihilations on nuclei.
Phys Lett B, 1988, {\bf 207}: 371--376.
\href{http://dx.doi.org/10.1016/0370-2693(88)90666-1}{DOI: 10.1016/0370-2693(88)90666-1}

\bibitem{Miyano:1984dc} Miyano K, Noguchi Y, Fukawa M, \emph{et al}.
Evaporation of Neutral Strange Particles in $\bar{p}$ Ta at 4 GeV/$c$.
Phys Rev Lett, 1984, {\bf 53}: 1725
\href{http://dx.doi.org/10.1103/PhysRevLett.53.1725}{DOI: 10.1103/PhysRevLett.53.1725}

\bibitem{Ahmad:1997fv} Ahmad S, Bonner B E, Buchanan J A, \emph{et al}. 
Strangeness production in antiproton nucleus interactions.
Nucl Phys B (Proc Suppl), 1997, {\bf 56A}: 118--121.
\href{http://dx.doi.org/10.1016/S0920-5632(97)00262-4}{DOI: 10.1016/S0920-5632(97)00262-4}

\bibitem{Gaitanos:2011fy} Gaitanos T, Larionov A B, Lenske H, Mosel U.
Formation of double-$\Lambda$ hypernuclei at PANDA.
Nucl Phys A, 2012, {\bf 881}: 240--254.
\href{http://dx.doi.org/10.1016/j.nuclphysa.2011.12.010}{DOI: 10.1016/j.nuclphysa.2011.12.010}

\bibitem{Gaitanos:2014bla} Gaitanos T, Lenske H.
Production of multi-strangeness hypernuclei and the YN-interaction.
Phys Lett B, 2014, {\bf 737}: 256--261.
\href{http://dx.doi.org/10.1016/j.physletb.2014.08.056}{DOI: 10.1016/j.physletb.2014.08.056}

\bibitem{Miyano:1988mq} Miyano K, Noguchi Y, Yoshimura Y, \emph{et al}. 
Neutral strange particle production and inelastic cross section in $\bar p$+Ta reaction at 4 GeV/c.
Phys Rev C, 1988, {\bf 38}: 2788--2798.
\href{http://dx.doi.org/10.1103/PhysRevC.38.2788}{DOI: 10.1103/PhysRevC.38.2788}

\end{thebibliography}

\end{document}